# Multiphoton Label-Free *ex-vivo* imaging using a custom-built dual-wavelength microscope with chromatic aberrations compensation

Andrea Filippi[1,2,3,7]*, Giulia Borile[1,3,7], Eleonora Dal Sasso[4,5], Laura Iop[4,5], Andrea Armani[5,6], Michele Gintoli[1,7], Marco Sandri[5,6], Gino Gerosa[4,5] and Filippo Romanato[1,3,7,8]

1 Department of Physics and Astronomy "G. Galilei", University of Padua, Via Marzolo 8, 35131 Padua, Italy.

2 Fondazione Bruno Kessler, Via Sommarive 18, 38123 Povo, Trento, Italy

3 Institute of Pediatric Research Città della Speranza, Corso Stati Uniti 4, 35127 Padua, Italy.

4 Department of Cardiac, Thoracic and Vascular Sciences, University of Padua, Via Giustiniani 2, 35128 Padua, Italy.

5 VIMM, Venetian Institute of Molecular Medicine, Via G. Orus 2, 35129 Padua, Italy.

6 Department of Biomedical Sciences, University of Padua, Via Ugo Bassi 58/B, 35131 Padua, Italy.

7 LaNN, Laboratory for Nanofabrication of Nanodevices, Corso Stati Uniti 4, 35127 Padua, Italy.

8 CNR-INFM TASC IOM National Laboratory, Area Science Park, S.S. 14 Km 163.5, 34012 Basovizza, Trieste, Italy.

* Correspondence to andrea.filippi.5@phd.unipd.it

# Abstract

Label-Free Multiphoton Microscopy is a very powerful optical microscopy that can be applied to study samples with no need for exogenous fluorescent probes, keeping the main benefits of a Multiphoton approach, like longer penetration depths and intrinsic optical sectioning, while opening the possibility of serial examinations with different kinds of techniques. Among the many variations of Label-Free MPM, Higher Harmonic Generation (HHG) is one of the most intriguing due to its generally low photo-toxicity, which enables the examination of specimens particularly susceptible to photo-damages. HHG and common Two-Photon Microscopy (TPM) are well-established techniques, routinely used in several research fields. However, they require a significant amount of fine-tuning in order to be fully exploited and, usually, the optimized conditions greatly differ, making them quite difficult to perform in parallel without any compromise on the extractable information. Here we present our custom-built Multiphoton microscope capable of performing simultaneously TPM and HHG without any kind of compromise on the results thanks to two, separate, individually optimized laser sources with full chromatic aberration compensation. We also apply our setup to the examination of a plethora of *ex vivo* samples in order to prove the significant advantages of our approach.

# (Introduction)

Multiphoton Microscopy (MPM) is a Laser-Scanning Microscopy technique based on a non-linear, strongly localized excitation of fluorescence. The "non-linear" nature of the process derives from the dependence of the absorption rate on a higher power of the illumination intensity. MPM does not have its optical sectioning capabilities severely limited by the fluorescence contributions from outside the depth of focus of the objective (as in Wide-Field Microscopy) and, therefore, represents one of the best non-invasive techniques to achieve imaging in deep explanted tissues or in living animals [1,2]. The most common MPM variation is the Two-Photon Microscopy (TPM), which takes advantage of the quasi-simultaneous absorption of two photons by a molecular receptor, in a single quantum event. This phenomenon was theoretically predicted by Maria Goeppert-Mayer in 1931 [3], but could be experimentally verified only after the advent of mode-locked lasers. Furthermore, the physical phenomenon of Two-Photon laser Excitation (TPE) laid dormant until Denk *et al.* devised a practicable Two-Photon Laser-Scanning fluorescence Microscope [4]. The TPE rate depends on the second power of the incident light intensity and is $\sim 10^{-14}$ times smaller than the Single-Photon Absorption rate, hence the successful implementation of TPM imaging requires very high photon fluxes, which in practice translates into the use of mode-locked laser sources with pulse durations below $1\ ps$ and frequencies of $\sim 100\ MHz$. Despite the need for very intense light sources, TPM presents several advantages over the classical single-photon techniques. For example, the wavelengths typically used in TPM are in the Near-IR, Mid-IR spectrum, making them less subjected to scattering or absorption from thick specimens, which translates into longer penetration depths. Furthermore, the TPE focal volume is very small due to the high photon flux required to achieve the Two-Photon Absorption [1], therefore TPM has an inherent capability of performing axial sectioning and shows reduced photo-bleaching (restricted to the focus plane) [4]. The lateral resolution of TPM scales (for high numerical aperture objectives) as:

$$R_{MP}(NA > 0.7) = \frac{0.383 \lambda_{ex}}{NA^{0.91}} \quad (1)$$

Where $\lambda_{ex}$ is the excitation wavelength and $NA$ is the objective's numerical aperture. Despite the obvious benefits of TPM, there are cases where it cannot (or should not) be applied. When the examined specimens are very susceptible to the illumination light, showing high rate of photo-bleaching and conspicuous photo-damages inside the focal volume, or when they will later undergo a series of other experimental examinations, and therefore must remain as much intact and unchanged from the original conditions as possible, the use of another variant of MPM, known as Higher Harmonic Generation (HHG), is much more indicated than TPM. HHG retains the intrinsic optical sectioning capabilities of TPM but, usually, does not involve the absorption of the excitation light, producing very little heat and resulting in a much less phototoxic technique than any other kind of microscopy based on fluorescence. Furthermore, HHG does not strictly require the use of molecular probes in order to generate and detect signal from a specific biological feature: any structure with characteristics that satisfy at least one order of harmonic generation can natively produce a detectable HHG signal, thus drastically reducing the specimen preparation requirements, induced alterations from the original conditions and maximizing its reusability in serial examinations. The most commonly used orders of harmonic generation are the Second Harmonic Generation (SHG) and Third Harmonic Generation (THG), which double and triple the incident light's frequency, respectively, and produce a narrow signal band of half and one-third the excitation wavelength. The SHG signal has a quadratic dependence on the input laser intensity and originates only from media without inversion symmetry (e.g. tissues made of chiral molecules [5-7]), while the THG signal has a cubic dependence on the input intensity (higher average powers required) but, in principle, can be elicited from any material to a varying degree. The resulting signals scale as follow [8]:

$$SHG \propto \frac{(p \cdot \chi_2)^2}{\tau} \quad ; \quad THG \propto \frac{(p \cdot \chi_3)^3}{\tau^2} \qquad (2)$$

where $p$ is the input power, $\chi_2$ and $\chi_3$ are susceptibility coefficients of the second and third order, respectively, and $\tau$ is the laser pulse width. From a microscopy practical point of view, usually a detectable THG signal originates from interfaces between media with a significant difference between their refractive indexes, like lipids immersed in aqueous fluids [9-11], cellular membranes [12,13] and protein aggregates [13,14]. However, both SHG and THG are very sensitive to external factors (e.g. temperature, medium ionic strength, pH) that can alter, even slightly, the molecular structures from which the signal originates [8]. Therefore, a careful optimization of the experimental parameters must be performed on sample basis in order to achieve the best possible HHG conditions. Here we report on the design and realization of a custom Multiphoton microscope capable of performing both TPM and HHG simultaneously through to the use of two separate laser sources that can be independently optimized for one technique or the other, in order to maximize the quality and quantity of the obtainable complementary information from different kind of biological tissues.

# Results

**Microscopy setup optimization.** First, we wanted to optimize the polarization state for both laser beams in order to achieve the best possible quality of the detectable signal. We positioned a polarization analyzer (Schafter&Kirchoff, Germany) under the scanning-head of the microscope to measure the polarization right before the back-aperture of the objective and we rotated a couple of $\lambda/2$ Wave Plate and $\lambda/4$ Wave Plate in order to set the polarization state of both lasers to circular. A circular polarization ensure the highest (on average) signal for the auto-fluorescence Multiphoton emission due to the generally random orientation of the intrinsic fluorophores' dipole axis [15] while, also, maximizing the SHG signal for randomly oriented collagen fibers. If the collagen structures of some samples happen to have a preferred orientation, the polarization can be easily set to linear and aligned to be at 45° from the fibers, which is the direction of generation of the maximum SHG signal [16]. A known issue of using a dual-wavelength excitation setup are the chromatic aberrations that can severely compromise the image quality. In order to overcome this problem we chose, as a starting point, an objective with a reported color-correction up to 1300 $nm$ and then we further corrected the small residual lateral and axial aberrations following a procedure commonly used in STED Microscopy, where perfect overlapping of two beams in the focal spot is a strictly required condition [17-19]. Therefore, the axial aberration was fine-tuned *via* a couple of telescopes (Thorlabs Inc.) that slightly changed the curvature radii of the two beams in order to move the focal spots along the optical axis and superimpose them with high precision, while the lateral displacement was corrected by superimposing the two Point Spread Functions (PSFs) *via* a silver mirror and a dichroic (Thorlabs Inc.)

mounted on piezo-controlled kinematic mirror mounts (Polaris, Thorlabs Inc.). In order to see the PSFs, we imaged a sample of Quantum Dots (diameter~70 $nm$, much smaller than the diffraction limited resolution) dispersed in a sol-gel $ZrO_2$ matrix. We checked the status of the axial and lateral displacements through the analysis of several z-stack of the beam's PSFs, using a custom-developed software (Figure 1). After careful fine-tuning, we achieved a full 3D PSFs superimposition, with a negligible residual lateral displacement and an axial displacement of $< 200\ nm$, near the minimum requirement for STED applications [20], which is a much higher resolution technique than simple Multiphoton Microscopy (3D-STED can achieve $< 100\ nm$ axial resolution, while MPM is diffraction limited to ~1 $\mu m$).

**Label-Free analysis of thin cryosections.** In order to test the quality of the obtainable images after the polarization calibration, we imaged mice heart cryosections with both lasers separately (Figure 2). The cryosections showed overall high signal intensity and very good S/N ratio for both auto-fluorescence emission and harmonic generation (SHG/THG). Many different features can be observed in cardiac cryosections without any labelling procedure. Remarkably, sarcomeres' myosin as well as collagen deposits surrounding the coronary vessel were clearly distinguishable due to their strong SHG signal from both 800 $nm$ and 1200 $nm$ excitation wavelengths. We evaluated the intensity profile of the SHG signal produced by the sarcomeres: it has a very good S/N ratio (Suppl. Figure S1) and, in line with previous reports [16], appears to arise from the myosin of thick filaments. Moreover, elastin can be visualized simultaneously by TPE. The excitation wavelength of 1200 $nm$ does not show some of the sample features that are clearly visible from the autofluorescence signal excited with 800 $nm$, instead, proving the dual-laser imaging approach more information-rich than the common single-wavelength TPM. In addition to intrinsic biological fluorophores, such as myosin, collagen and elastin, that report solely on structural information, nicotinamide adenine dinucleotide (NADH) emits a strong TPE signal [21] that is related to the metabolic state of the imaged cells. We thus applied our setup to study skeletal muscle cryosections, in order to discriminate glycolytic and oxidative muscle fibers depending on their NADH/mitochondria content [22]. As shown in Figure 3, two consecutive gastrocnemius muscle cryosections display fibers with higher TPE signal corresponding to those with higher SDH intensity in standard histology, consistently with the NADH origin of the TPE auto-fluorescence. Oxidative fibers show stronger SDH activity upon staining, corresponding to fibers having larger number of mitochondria, more oxidative capacity and smaller cross-sectional area. Upon repeated investigations of several cryosections (Figure 3-D), we confirmed the strict correlation between TPE and SDH intensity, thus validating the use of Label-Free TPE as a possible candidate for NADH quantification studies. Furthermore, we confirmed that more oxidative fibers having stronger TPE emission are those with smaller cross-sectional area (Suppl. Figure S2). Moreover, from the same cryosection we obtained information on collagen content (*via* SHG signal) without any need of additional staining processes. We performed a Hematoxylin/Eosin staining in the consecutive cryosection of the same muscle (Figure 3-C) in order to verify that the SHG signal really gives the same information about collagen as a conventional, widely used, staining method.

**Label-Free analysis of thick *ex-vivo* samples.** We then wanted to ensure that the excellent optimization work resulted in good quality images also in a more *in vivo* scenario; therefore, we proceeded to examine freshly excised unstained mouse lungs, as a good example of challenging thick sample. As can be seen in Figure 4, the SHG signal made possible to see very distinctly the collagen fibrillary structure of the lungs while the green auto-fluorescence of the intrinsic fluorophores highlighted the alveoli inside the lungs, as well as the nuclei of several cells scattered inside the lung tissue. Both collagen features and cell's nuclei remained perfectly visible and crisp even at relatively high zoom level, without any sign of photo-bleaching or photo-damages, confirming the power of this Label-Free technique when properly optimized.

**Dual-wavelength Label-Free imaging of bovine pericardium.** The main advantage of using a microscopy setup with two independent laser sources is that we can achieve the optimal illumination conditions for two distinct features without any compromise on signal strength or excitation efficiency. This is particularly useful for a Label-Free imaging technique, where, in general, the detected signals are significantly lower than the average fluorescence signal coming from commonly used fluorophores and the requirements needed to obtain a good S/N ratio are more stringent. In order to show what our custom build setup can achieve in that sense inside a biologically relevant Label-Free environment, we imaged a portion of decellularized bovine pericardium using both laser sources simultaneously (Figure 5). The sample was primarily made of collagen fibrils; therefore, one of the laser sources was entirely dedicated to optimize the SHG emission from the bundles. Another significant part of the sample consisted of elastin, which has its auto-fluorescence excitation maximum at $425\ nm$ [23], thus requiring a two-photon excitation beam tuned around $800\ nm$ in order to achieve the highest emission signal. However, collagen shows a significant absorption of blue photons [23] which competes with the two-photon excitation of the elastin's auto-fluorescence and the detection of the SHG signal coming from the collagen itself, due to auto-absorption, resulting in a lower S/N ratio at a given excitation power and a lower penetration depth for the SHG. We solved this problem tuning one laser source to $800\ nm$ and the other to $1200\ nm$: the first beam was optimized for the imaging of elastin's auto-fluorescence while the second beam was optimized for the SHG of the collagen. At $1200\ nm$, no significant elastin excitation was measured, the SHG signal was significantly stronger, due to lack of competing processes, while still be clearly distinguishable from the auto-fluorescence and the achieved penetration inside the tissue was of a few hundreds of microns, thus enabling 3D reconstruction studies on optically thick samples. Z-stacks performed on thick tissue showed that the SHG signal could be detected without any problem up to 200 $\mu m$ deep in the sample (Suppl. Movie M1).

# Discussion

We have presented our custom build Multiphoton microscope with two laser sources, capable of performing TPM and HHG simultaneously with the possibility of fine-tuning a specific source for one or the other. Our approach combines the best traits of both imaging techniques (e.g. intrinsic optical sectioning, reduced photo-bleaching, Label-Free capabilities), while avoiding sacrificing image quality and signal strength due to a necessary compromise on the selection of a single excitation wavelength. This is particularly relevant for HHG, where the signal intensity is extremely sensitive to many external factors and the need of a careful optimization of the experimental parameters on sample basis is stringent. We show that after a calibration of the laser's polarization states the microscope can acquire images with a very high signal intensity and good S/N ratio for both TPE and HHG while enabling simultaneous Label-Free visualization of different structures on which quantification studies can be successfully performed in order to obtain biologically relevant information without significantly altering the samples. We also prove that all the capabilities of our setup remain intact while studying thick *ex vivo* samples, freshly excised and totally Label-Free, detecting significantly strong signals up to 200 $\mu m$ deep in the sample. We fully take advantage of the dual laser configuration while examining decellularized bovine pericardium, achieving optical penetration inside the tissue of hundreds of micron and performing 3D reconstructions from z-stacks of optically thick samples without the need of any molecular probe labelling. The use of Label-Free techniques preserve the samples as similar as possible to their original conditions even after several hours and many imaging sessions, maximizing the repeatability and the ability of performing different examinations sequentially. Finally, we want to emphasize the complete customizability of our dual laser setup, which can be easily adapted to any specific needs for peculiar samples or can be upgraded to perform more complex Label-Free imaging, like Coherent Anti-Stokes Raman Spectroscopy (CARS) and Stimulated Raman Spectroscopy (SRS) Microscopy, or even super-resolution nanoscopy. The possibility of expanding our setup in order to achieve Two-Photon Excitation STimulated Emission Depletion (STED) Nanoscopy is currently being examined.

# Methods

**Multiphoton microscope.** The main laser source is a mode-locked Ti:Sapphire pulsed laser (Chameleon Ultra 2, Coherent), with pulses of ~140 $fs$ at 80 $MHz$ and tunable emission wavelength of $700-900\ nm$, which also serves as a pump for the second laser source: an Optical Parametric Oscillator (Compact OPO,

Coherent) with tunable emission wavelength in the $1000-1550\ nm$ range. Both lasers pass through a pair of Pockels Cells, voltage-controlled wave plates that combined with a polarizer can modulate the laser beam power with KHz frequency (ConOptics Inc.). A couple of silver-coated mirrors (Thorlabs Inc.) deflect them to the recombination area, where they are spatially overlapped by means of another silver mirror and a dichroic (Thorlabs Inc.) mounted on piezo-controlled kinematic mirror mounts (Polaris, Thorlabs Inc.). From the recombination area, the beams enter the scanning head (Bergamo Series, Thorlabs Inc.), equipped with a scan lens-tube lens and a Galvo/Resonant scanner ($8\ KHz$ scanning rate) and arrive on the sample through a water-immersion objective (Olympus XLPLN25XWMP2). The polarization state of both laser beams can be controlled independently by two achromatic $\lambda/4$ Wave Plates (Thorlabs Inc.) preceded by two achromatic $\lambda/2$ Wave Plates (Thorlabs, Inc.), which can compensate the polarization variations induced by the optical components in the scanning head in order to achieve the desired polarization state before the objective back-aperture with an high grade of purity. The light scattered or emitted from the sample is collected in epi-direction and, thanks to a longpass $705\ nm$ dichroic mirror (Thorlabs Inc.), is diverted to the detection module. This module consists of 4x GaAsP PMT detectors (H7422-40, Hamamatsu), with bandpass filters ($395/25\ nm$ Chroma, $460/50\ nm$ Chroma, $525/40\ nm$ Semrock, $625/90\ nm$ Semrock) separated, in terms of spectrum wavelength detection, by 3x dichroic mirrors ($425\ nm$ Chroma, $495\ nm$ Chroma, $565\ nm$ Chroma). A schematic representation of our setup can be seen in Suppl. Figure S3. The microscope can also perform trans-illumination experiments. In this case, an InGaAs detector (DET08CL $800-1800\ nm$, Thorlabs Inc.) is positioned near the LED illumination source below the sample holder. A longpass dichroic mirror (Thorlabs Inc.) separates the transmitted illumination path from the transmitted detection path.

**Heart cryosections.** Mice were sacrificed by cervical dislocation and hearts were quickly harvested and cut in two portions in the transverse direction and processed as described in [24]. Briefly, blood clots were carefully removed the heart as fixed to maintain structural integrity with 1% paraformaldehyde in phosphate buffered saline (PBS 1X: 137 $NaCl$, 2.7 $KCl$, 10 $Na_2HPO_4$, 1.8 $KH_2PO_4$, in mM) at room temperature for 15 minutes. After 3 washes of 5 minutes with PBS 1X, hearts were allowed to dehydrate in sucrose 30% (w/v in distilled water) at $4°C$ overnight. The following day, hearts were embedded in OCT freezing medium (Optimal Cutting Temperature, Kaltec) and carefully frozen in liquid nitrogen vapor. Frozen samples were maintained at $-80°C$. Frozen hearts were cut in $10\ \mu m$ slices using a cryostat (Leica CM1850, Leica Microsystems GmbH, Wetzlar, Germany) and placed on superfrost glass slides (Vetrotecnica) maintained at $-80°C$ until use.

**Gastrocnemius muscle and SDH staining.** Gastrocnemius muscles were harvested from mice and immediately frozen in liquid nitrogen. Ten-micron cryosections were obtained with a cryostat (Leica CM1850, Leica Microsystems GmbH, Wetzlar, Germany) and stained for Succinate dehydrogenase (SDH). Processed cryosections were examined in a fluorescence microscope (Olympus BX60), as described in [25].

*Ex-vivo* **lungs.** Mouse was sacrificed by cervical dislocation and lungs were quickly harvested and carefully washed in ice-cold PBS to avoid blood clotting. Lungs were then transferred on a petri dish filled with PBS. To prevent curling and movement, lungs were held down through a homemade platinum holder.

**Decellularized bovine pericardium.** Bovine pericardia were collected from the local slaughterhouse and decellularized using a method based on alternated hypo- and hypertonic solutions, detergents (Triton X-100 and sodium cholate, Sigma-Aldrich, Saint Louis, MO, USA) and non-specific endonucleases (Benzonase, Sigma-Aldrich) [26,27]. Following the decellularization procedure, samples of $1\ cm^2$ were placed into plastic embedding devices (Bio Optica, Milano, Italy) and covered by a thin layer of 4% low melting agarose solution prepared in PBS (Sigma-Aldrich).

**Statistical analysis.** For correlation analysis, images were processed with Fiji using the ROI Manager plug-in in order to obtain mean ROI intensity and Area values. Pearson correlation coefficient was evaluated with OriginPro™ 2016 and data were fitted with linear regression.

**Data availability.** The datasets generated during and/or analyzed during the current study are available from the corresponding author on reasonable request.

# Acknowledgements

Author AF is supported by *Fondazione Bruno Kessler* (FBK) PhD fellowship. Author GB is supported by University of Padua (Bando per il Finanziamento di Assegni Dipartimentali, Bando 2016). The authors are grateful to Prof. Marco Mongillo for sharing the heart cryosection samples.

# Author Contributions

AF and GB wrote the main manuscript text, prepared figures and supplementary information. AF acquired and processed TPE and HHG images. MG analyzed PSFs for the displacements correction. EDS and LI prepared decellularized tissue sample, related methods section and contributed to image acquisition. AA prepared skeletal muscle samples and performed SDH staining. GB prepared heart cryosections and lung tissue, performed image analysis and Hematoxylin/eosin staining. FR, GG and MS contributed to results interpretation. All authors reviewed the manuscript.

## Competing Financial Interests

The authors declare no conflicting financial interests.

# Figures

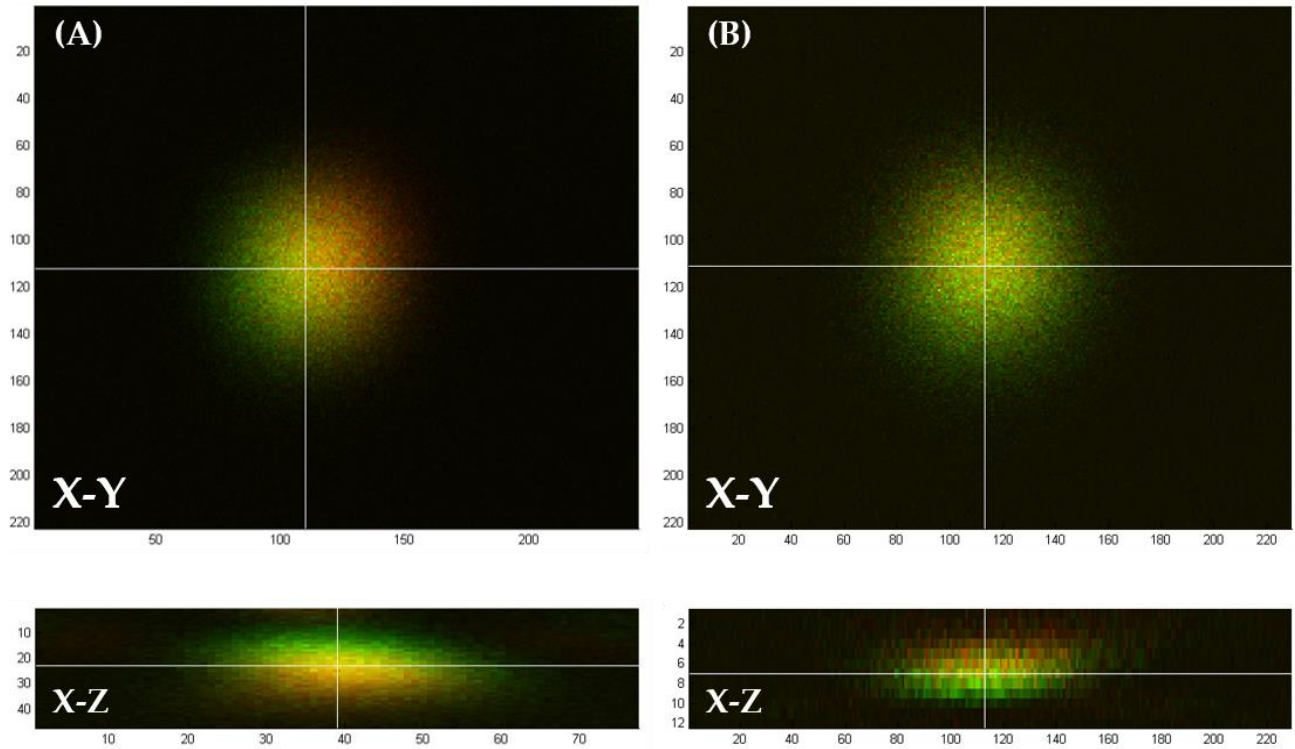

**Figure 1. PSFs displacement comparison before and after correction.** **(A)** PSFs displacement in the X-Y (top) and X-Z (bottom) planes before fine-tuning. **(B)** PSFs displacement in the X-Y (top) and X-Z (bottom) planes after fine-tuning. The final lateral displacement is negligible while the residual axial displacement is < 200 $nm$, well below the minimum requirements for simple Multiphoton Microscopy (note the difference in scales between (A) and (B) X-Z planes). Axis scales express coordinates in pixels (4.17 $nm/pixel$ X and Y, 100 $nm/pixel$ Z).

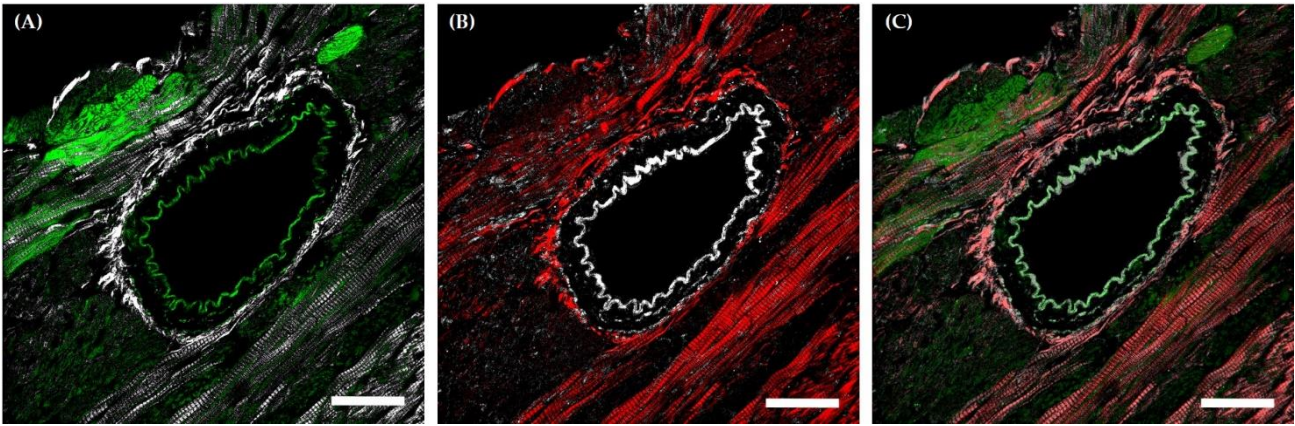

**Figure 2. Simultaneous TPE and HHG images of cardiac cryosections.** **(A)** Cardiac cryosection imaged with $800\ nm$ laser wavelength shows TPE auto-fluorescence (Green, $\lambda_{em} = \sim 500\ nm$) and SHG signal (White, $\lambda_{em} = 400\ nm$). **(B)** Same field of view imaged with $1200\ nm$ laser wavelength shows SHG (Red, $\lambda_{em} = 600\ nm$) and THG (White, $\lambda_{em} = 400\ nm$). **(C)** Superposition of the same field imaged with $800\ nm$ and $1200\ nm$; The SHG produced by the two different laser beams, as well as the elastin autofluorescence and its THG signal, perfectly superimpose, testifying the excellent spatial alignment of the lasers. The auto-fluorescence excited with $800\ nm$ also highlight some features that are not visible with an excitation wavelength of $1200\ nm$, proving the dual-laser imaging approach more information-rich than the common single-wavelength TPM. Scale bars 50 µ$m$.

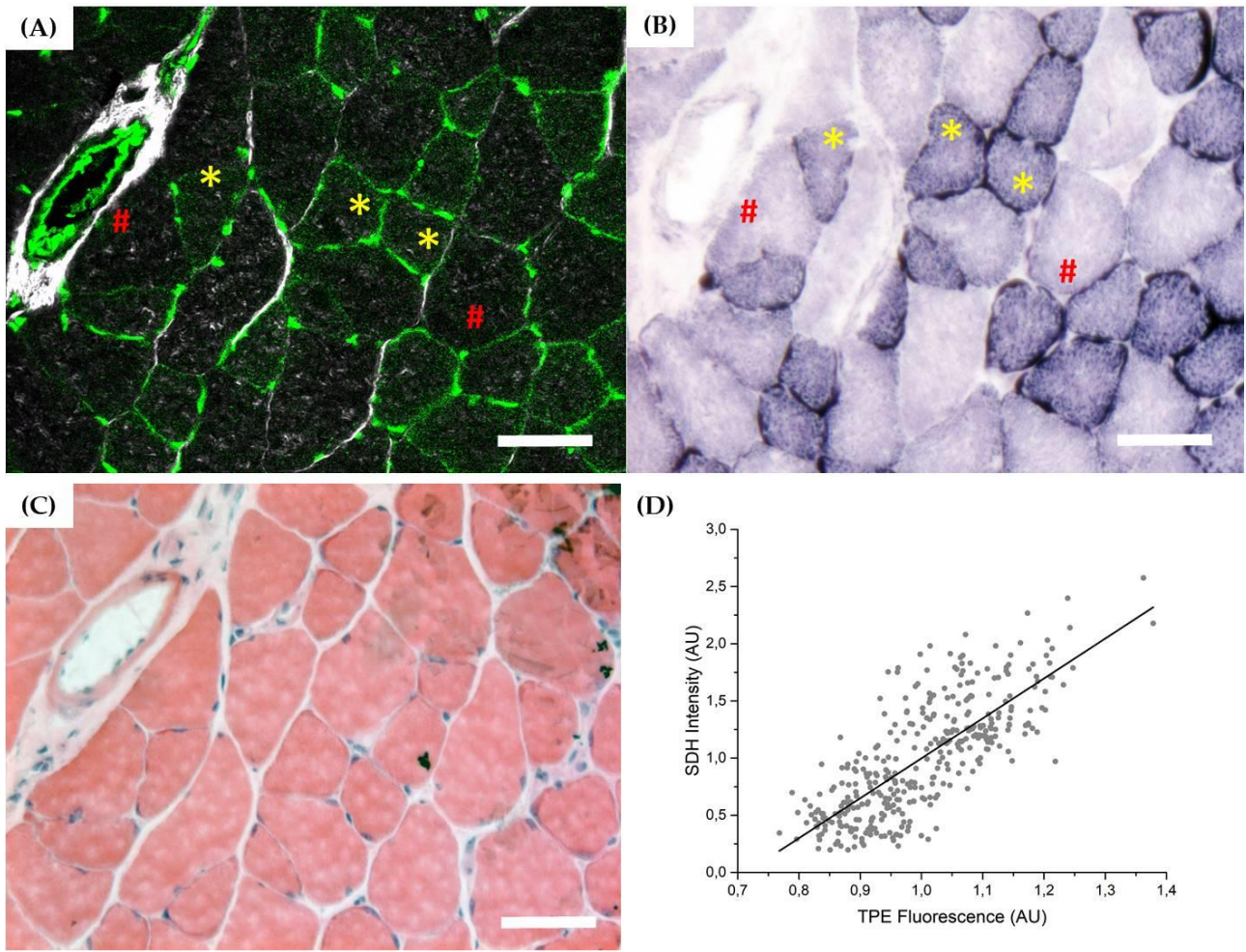

**Figure 3. TPE of NADH in muscle cryosections.** **(A)** TPE (Green, $\lambda_{em} = \sim500\,nm$) and SHG (White, $\lambda_{em} = 400\,nm$) signals from gastrocnemius muscles cryosection. **(B)** SDH staining of the consecutive cryosection from the same muscle. **(C)** Hematoxylin/Eosin staining of the consecutive cryosection from the same muscle. **(D)** Correlation analysis of TPE and SDH signal intensity from the same fiber ($N = 4$ cryosections, $n = 350$ fibers). Pearson correlation coefficient $P > 0.75$. Yellow (*) and red (#) highlight corresponding fibers, oxidative and glycolytic respectively. Scale bars 100 µ$m$.

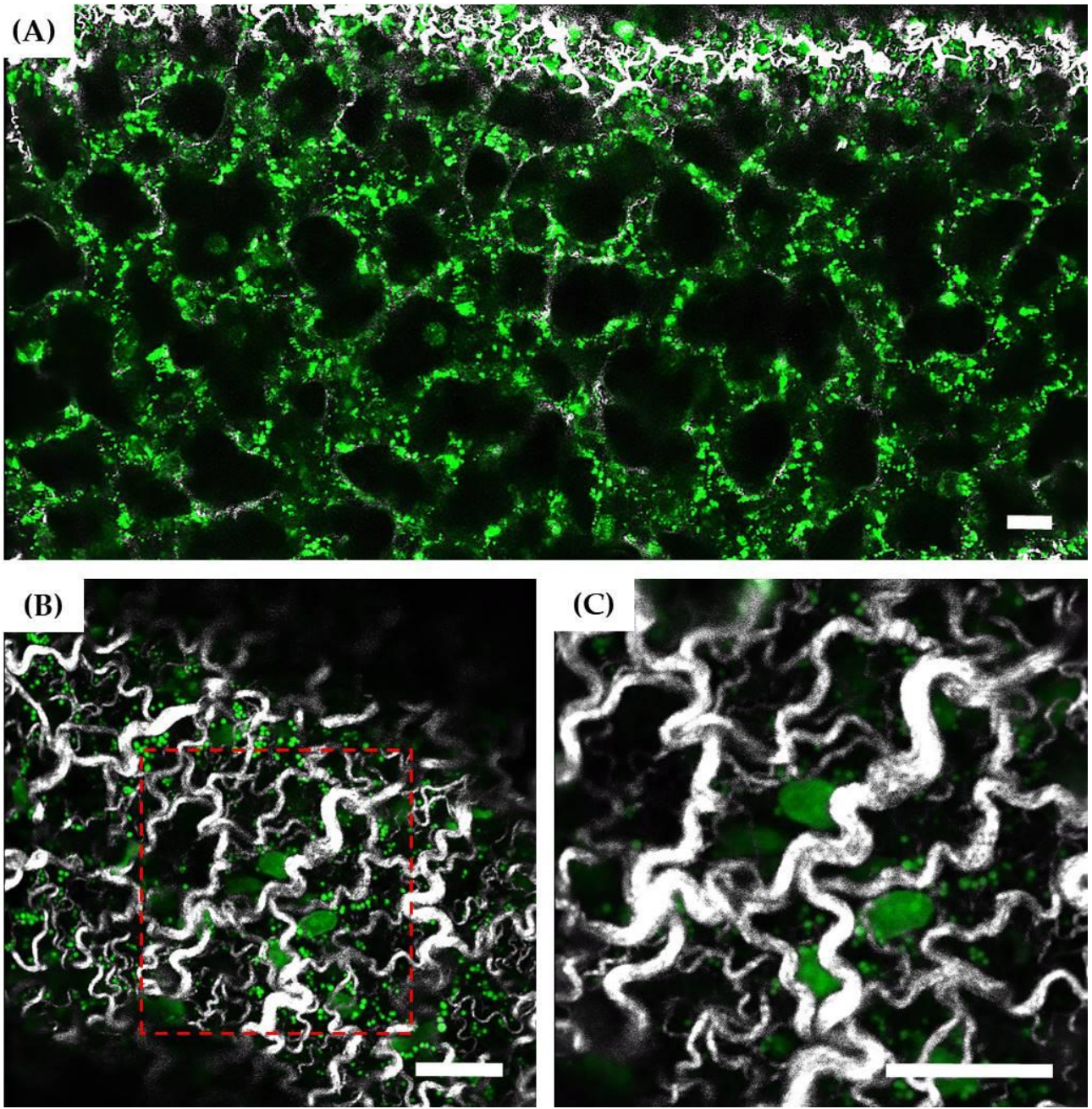

**Figure 4. Simultaneous TPE and SHG images of ex-vivo lung tissue. (A)** TPE and SHG signals imaged with $800\,nm$ wavelength. Alveolar sacs can be recognized thanks to their TPE auto-fluorescence emission (Green, $\lambda_{em} = \sim 500\,nm$) and are surrounded by the coarse collagen fibers of the serosa (White, $\lambda_{em} = 400\,nm$). **(B)** and **(C)** are two magnifications of the serosa collagen fibers. Scale bars $20\,\mu m$.

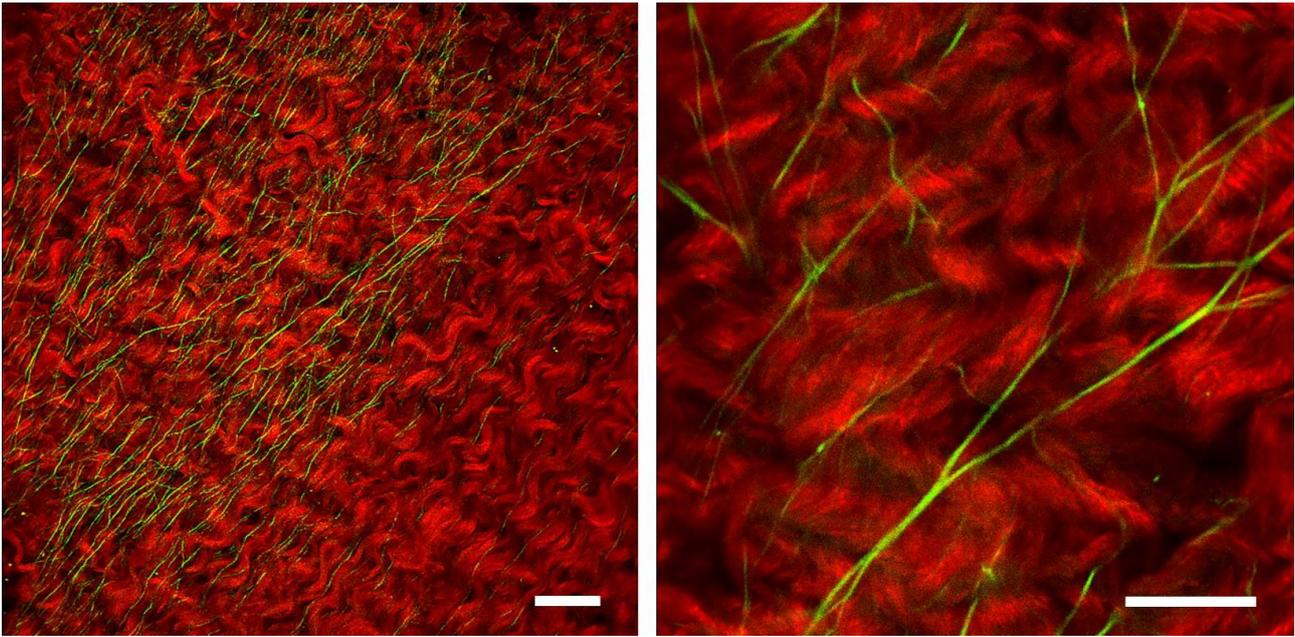

**Figure 5. Simultaneous TPE and SHG images of decellularized pericardium for scaffold characterization.** SHG signal from collagen bundles imaged with $1200\ nm$ wavelength (Red, $\lambda_{em} = 600\ nm$) and elastin autofluorescence excited with $800\ nm$ wavelength (Green, $\lambda_{em} = \sim500\ nm$). Scale bars $50\ \mu m$.

# Multiphoton Label-Free *ex-vivo* imaging using a custom-built dual-wavelength microscope with chromatic aberrations compensation

Andrea Filippi[1,2,3,7]*, Giulia Borile[1,3,7], Eleonora Dal Sasso[4,5], Laura Iop[4,5], Andrea Armani[5,6], Michele Gintoli[1,7], Marco Sandri[5,6], Gino Gerosa[4,5] and Filippo Romanato[1,3,7,8]

# Supplementary Information

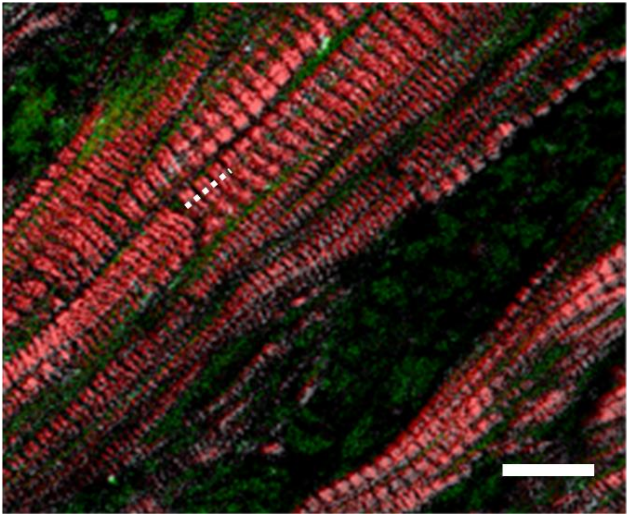 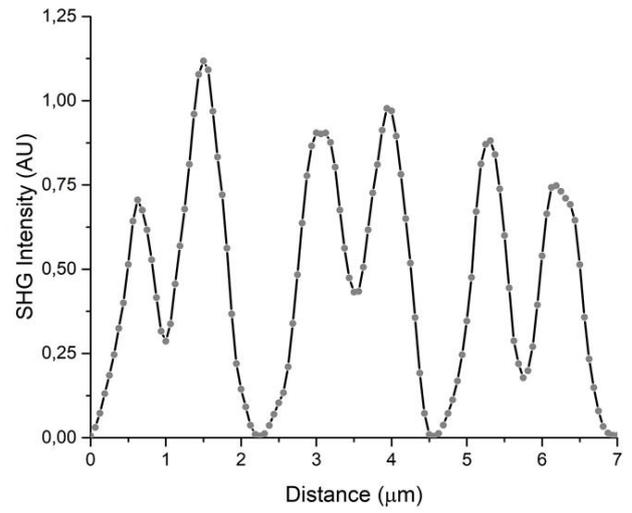

**Supplementary Figure S1. Analysis of SHG signal from thick filament of the sarcomeres.** (A) Detail of Figure 1 (C). (B) SHG normalized intensity profile of the dotted line in (A) shows SHG-bright double-bands clearly visible thanks to the good signal-to –noise ratio. Scale bar 10 $\mu m$.

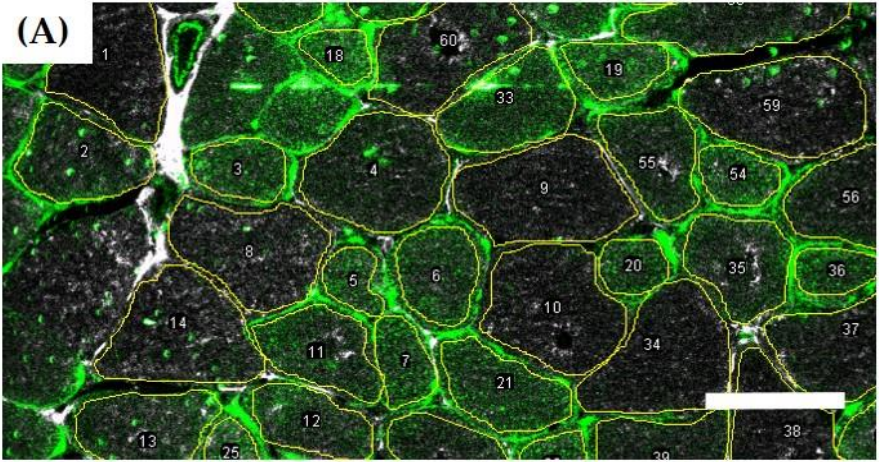

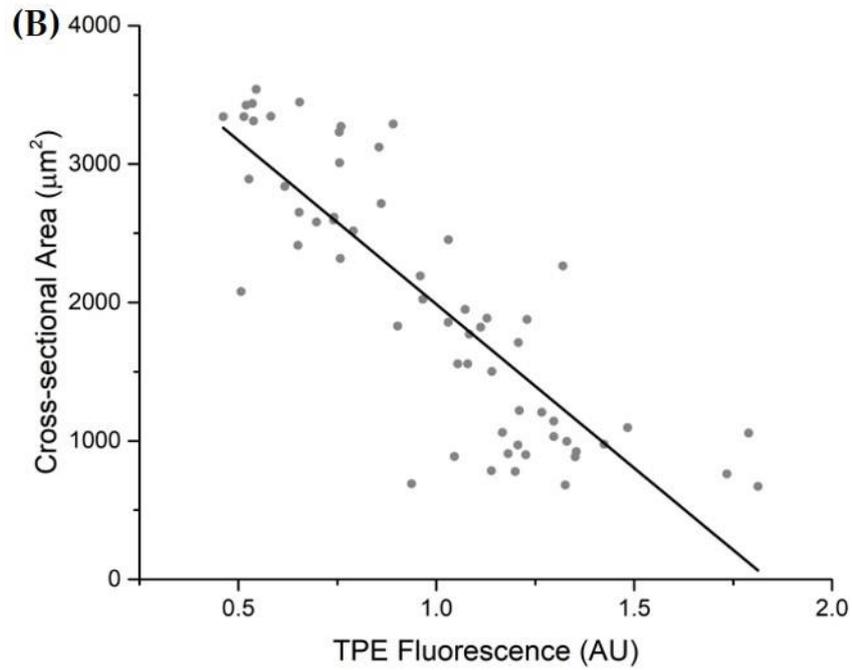

**Supplementary Figure S2. Correlation between fiber cross-sectional area and TPE intensity.** **(A)** ROI analysis of fibers within a muscle cryosection. **(B)** Cross-sectional area plotted against TPE fluorescence shows that smaller fibers are richer in mitochondria content. Pearson correlation coefficient: $P > 0.85$. Scale bar 50 $\mu m$.

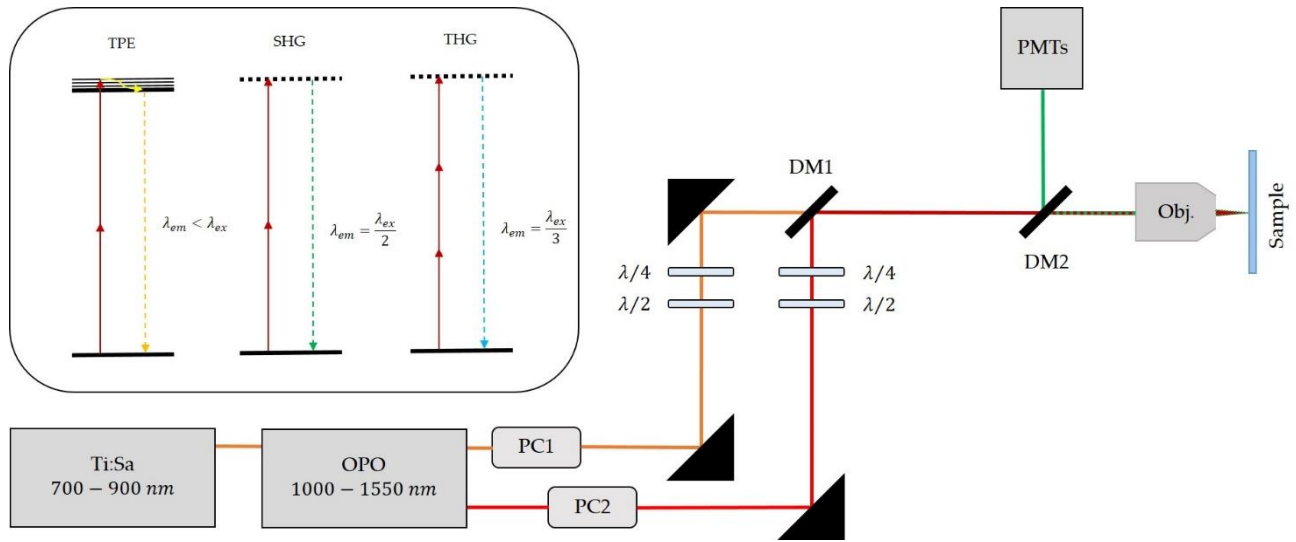

**Supplementary Figure S3. Schematic representation of the setup.** Both laser beams coming from the Ti:Sa laser and the OPO are modulated in intensity by two Pockels Cells (PC1,PC2), circularly polarized by a couple of Half-Wave Plates and Quarter-Wave Plates ($\lambda/2$, $\lambda/4$), spatially overlapped by means of a Dichroic Mirror (DM1) and finally focused on the sample thorough a 25x Objective. The fluorescence and HHG signals are detected in epi-direction and sent to four GaAsP PMT detectors thanks to another Dichroic Mirror inside the scanning-head (DM2).

(See MovieM1)

**Supplementary Movie M1. Three-dimensional reconstruction of decellularized bovine myocardium.** Z-stack of 200 $\mu m$ in 20 steps reconstructed with Fiji. Scale bar 50 $\mu m$.